\documentstyle[aps,multicol,epsf]{revtex}
\def\be{\begin{equation}}
\def\bea{\begin{eqnarray}}
\def\bma{\begin{mathletters}}
\def\ee{\end{equation}}
\def\eea{\end{eqnarray}}
\def\ema{\end{mathletters}}

\begin{document}
\author{Vlatko Vedral \footnote{e-mail:
v.vedral@ic.ac.uk}}
\address{Optics Section, The Blackett Laboratory, Imperial College, London SW7 2BZ, United Kingdom}
\title{Entanglement in The Second Quantization Formalism}
\date{\today}

\maketitle

\begin{abstract}
We study properties of entangled systems in the (mainly non-relativistic) second
quantization formalism. This is then applied to interacting and non-interacting bosons
and fermions and the differences between the two are discussed. We present a general
formalism to show how entanglement changes with the change of modes of the system.
This is illustrated with examples such as the Bose condensation and the Unruh effect.
It is then shown that a non-interacting collection of fermions at zero temperature can
be entangled in spin providing that their distances do not exceed the inverse Fermi
wavenumber. Beyond this distance all bipartite entanglement vanishes, although
classical correlations still persist. We compute the entanglement of formation as well
as the mutual information for two spin-correlated electrons as a function of their
distance. The analogous non-interacting collection of bosons displays no entanglement
in the internal degrees of freedom. We show how to generalize our analysis of the
entanglement in the internal degrees of freedom to an arbitrary number of particles.
\end{abstract}

\section{Introduction}

Entanglement is a phenomenon that has become of great importance in modern
applications of quantum mechanics \cite{Vedral}. We usually think of entanglement as
existing between different degrees of freedom of two or more particles. There has been
a number of important advances recently in understanding entanglement in systems
containing a small number of particles. If we are to use this appropriately, however,
we need to extend our analysis to realistic systems containing a large number of
particles. When it comes to large systems in quantum mechanics, the concept of a
particle actually fades away and is replaced by the notion of ``an excitation of a
given mode of the field representing the particle". Individual particles actually
become indistinguishable (this is, of course, also true for a small number of
identical particles, but indistinguishability may not play any significant role). In
addition to that, the concepts of different particle statistics (fermions versus
bosons) then also becomes directly relevant and it is important to understand its
relation to entanglement. The most convenient and appropriate formalism to deal with
all the issues involving a large number of systems is the second quantization. So we
need to understand what entanglement means in this setting if we are to be able to
harness solid state and condensed matter systems for information processing purposes.
Another benefit of the second quantization is that it is also the correct formalism
for the (relativistic) quantum field theory. We can therefore exploit some already
known results in this field in the hope of reaching a more complete understanding of
entanglement.

In this paper we will see that entanglement is highly dependent on the choice of modes
\cite{Zanardi} (for an alternative treatment see \cite{Schliemann}). The choice of
modes is mainly dictated by the physics of the given situation, and the general rule
is that in order for modes to be entangled they need to interact in some way. The crux
is that what happens to be an interacting hamiltonian for one set of modes may not be
so for a different set of modes. Alternatively stated, the vacuum state of one set of
modes, may not be the vacuum state when the modes are changed. It is this simple fact
that will be discussed in this paper and shown to be at the root of many different
phenomena. More interestingly, however, we will see that the modes need not interact
in order to produce entanglement - entanglement can be a consequence of
indistinguishability of quantum particles and the existence of quantum statistics.

In the next section we review the formalism of second quantization. In section $3$ we
present a very simple illustrative example of two harmonic oscillators and show how
entanglement arises through their interaction. The conclusions of this section are
very general and will then be applied to the Bose condensation as well as the Unruh
effect in section $4$. In section $5$ we show that interaction is, however, not
necessary to generate entanglement. We present a Fermi gas of non-interacting
particles at zero temperature and show that the spin degree of freedom of individual
fermions can be entangled even though the fermions themselves are non-interacting. We
show that an analogous collection of non-interacting bosons, on the other hand, cannot
generate any such entanglement in the internal degrees of freedom. We finally conclude
and discuss some future directions in section $6$.

\section{Second Quantization}

First we briefly review the formalism of the second quantization \cite{Weinberg}. We
will approach the second quantization formalism from the perspective of many particle
systems, rather than as a means of achieving relativistic quantum theory. Suppose we
have the Schr\"odinger equation for a single quantum system in one dimension (for
simplicity):
\begin{equation}
\hat H(x) \Psi (x,t) = i \hbar \frac{\partial}{\partial t} \Psi
(x,t)
\end{equation}
Note that although this will lead to the non-relativistic second
quantization, much of what we say will be true relativistically as
well. A formal solution to this equation is
%s
\begin{equation}
\Psi (x,t) = \sum_n b_n (t) \psi_n (x)\; .
\end{equation}
As usual, by substituting this back into the Schr\"odinger
equation, we obtain its time independent version
\begin{equation}
\hat H (x) \psi_n (x) = E_n \psi_n (x) \; ,
\end{equation}
where $E_n$ are the corresponding energies. In order to convert
this into a many-particle equation, we apply the formalism of
second quantization, which means that we effectively have to
``upgrade" the wave-function $\Psi$ into an operator. Formally, we
write
\begin{equation}
\hat \Psi (x,t) = \sum_n \hat b_n (t) \psi_n (x)
\label{xtop}
\end{equation}
The $\hat b_n$ operators are the well-known annihilation
operators. The conjugate of the above equation becomes
\begin{equation}
\hat \Psi^{\dagger} (x,t) = \sum_n \hat b^{\dagger}_n (t)
\psi_n^{*} (x)
\end{equation}
where $b^{\dagger}_n$ are the creation operators. In this paper we
only consider the time independent creation and annihilation field
operators. Nothing much would change conceptually if the field
operators were to be time dependent, although mathematically the
whole analysis would become much more involved. The Hamiltonian
also has to be second quantized, and the new Hamiltonian is given
by the average of the old (first quantized) Hamiltonian
\begin{equation}
\tilde H = \int dx \hat \Psi^{\dagger} (x,t) \hat H (x) \hat \Psi
(x,t)
\end{equation}
By invoking the orthogonality rules, $\langle \psi_n |\psi_m
\rangle = \delta_{mn}$, the second quatized Hamiltonian becomes
\begin{equation}
\tilde H = \sum_n E_n \hat b_n^{\dagger} \hat b_n
\end{equation}
and this is the same as a set of independent harmonic oscillators. Now, the fields
here are clearly non-interacting and therefore we may conclude that the $b$ modes are
disentangled. But, we have to be careful. While there is, clearly, no entanglement in
the $b$ modes, there may be entanglement between some other modes. To illustrate this
let us have a look at the simplest case of two harmonic oscillators. Note that the
index $n$ can also contain the internal degrees of freedom (such as spin or
polarization), however, not until section $5$ will we need to take this complication
into account.

\section{Interacting Harmonic Oscillators}

Suppose that we have two non-interacting harmonic oscillators so
that the total Hamiltonian is (we will omit hats from now on):
\begin{equation}
H = \hbar (\omega - \lambda) b_1^{\dagger}b_1 + \hbar (\omega +
\lambda) b_2^{\dagger}b_2
\end{equation}
Since oscillators are non-interacting their eigenstates are, therefore, composed of
disentangled direct products of the eigenstates of the individual harmonic
oscillators. For example, the ground state is just a product of the ground states of
the oscillator $1$ and the oscillator $2$, $|0\rangle_b = |0\rangle_{b_1} \otimes
|0\rangle_{b_2}$, so that it is annihilated by both the annihilation operators of
individual oscillators
\begin{eqnarray}
b_1 |0\rangle_b & = & 0\\
b_2 |0\rangle_b & = & 0 \; .
\end{eqnarray}
However, let us make a change of modes and see what happens to the
new eigenstates. Let the new creation and annihilation operators
be
\begin{eqnarray}
a_1 = \frac{b_1 + b^{\dagger}_2}{\sqrt{2}} \\
a_2 = \frac{b^{\dagger}_1 - b_2}{\sqrt{2}}
\end{eqnarray}
(We could have made a more general change, but the one above is
sufficient to illustrate the point here). The original Hamiltonian
now becomes
\begin{equation}
H = \hbar \omega (a_1^{\dagger}a_1 + a_2^{\dagger}a_2) + \hbar
\lambda (a_1^{\dagger}a_2 + a_2^{\dagger}a_1)
\end{equation}
and it is transparent that the $a$ modes actually represent two
coupled harmonic oscillators. It is, therefore, not surprising
that the ground state of this system is actually entangled when
written in the basis of the original harmonic oscillators. It is
given by (up to a normalization which is not relevant for our
discussion) \cite{Han}
\begin{equation}
|0\rangle_a \propto \sum_n [-i\tanh \frac{4\lambda}{\omega}]^n |n\rangle_{b_1}\otimes
|n\rangle_{b_2} \; ,
\end{equation}
where $|n\rangle_{b_i}$ is the state containing $n$ quanta in the oscillator $b_i$.
This state is clearly entangled with respect to eigenstates of $b$ s, and it is
written in the Schmidt decomposition form making it easy to detect entanglement. The
amount of entanglement is obtained by tracing out one oscillator and calculating the
entropy of the resulting mixture \cite{Vedral}, which is known as the entropy of
entanglement. The resulting entanglement is
\begin{equation}
E =  \frac{2\ln (\coth \frac{4\lambda}{\omega})}{\frac{1}{\tanh^2
\frac{4\lambda}{\omega}} - 1} - \ln (1 - \tanh^2 \frac{4\lambda}{\omega})\; .
\end{equation}
To make the result look more familiar to us (and to prepare
ourselves for the next section) we introduce the following change
of variables
\begin{equation}
[\tanh \frac{4\lambda}{\omega}]^2 = e^{-\frac{\hbar \omega}{kT}},
\end{equation}
where we think of $T$ as the corresponding (effective)
temperature; the entanglement is now given by
\begin{equation}
E =  \frac{\hbar \omega}{kT}\frac{1}{e^{\frac{\hbar \omega}{kT}} -
1} - \ln (1 - e^{-\frac{\hbar \omega}{kT}}) \label{entharm}
\end{equation}
which is just the entropy of a thermal state of a harmonic
oscillator, i.e. it is the familiar Planck distribution. We see
that if $T\rightarrow 0$ (i.e. the interaction tends to zero,
$\lambda \rightarrow 0$), then $E\rightarrow 0$, and entanglement
disappears as expected. The fact that this is a thermal spectrum
features strongly in the Unruh effect and the Hawking radiation
from a black hole \cite{Davies} as will be seen in the next
section.

Let us now generalize our consideration to an arbitrary field which can be written as
superposition of the creation and annihilations operators for the various modes
\begin{equation}
\psi (x) = \sum_n (\psi_n(x) a_n + \psi^*_n(x) a^{\dagger}_n) \; .
\end{equation}
Here we basically have infinitely many harmonic oscillators to sum
over. The vacuum state for the $a$ operators is defined as $a_n
|0\rangle = 0$ for all $n$. Now, suppose that we expand the field
in terms of a different set of creation and annihilation
operators, like so
\begin{equation}
\psi (x) = \sum_n (\tilde \psi_n(x) b_n + \tilde \psi^*_n(x)
b^{\dagger}_n) \; .
\end{equation}
The new operators can always be expanded as a combination of the
old operators as
\begin{eqnarray}
b_m & = & \sum_n \alpha_{mn} a_n + \beta_{mn} a^{\dagger}_n \\
a_m & = & \sum_n \alpha^*_{mn} b_n - \beta^*_{mn} b^{\dagger}_n
\end{eqnarray}
(the new mode functions $\tilde \psi$ can also be expressed in
terms of the old ones \cite{Davies}, but this is not important for
our discussion). Due to the unitarity requirement, the
coefficients of transformation have to posses the following
properties:
\begin{eqnarray}
& \sum_k & (\alpha_{ik}\alpha^*_{jk} - \beta_{ik}\beta^*_{jk}) = \delta_{ij} \\
& \sum_k & (\alpha_{ik}\beta_{jk} - \beta_{ik}\alpha_{jk}) = 0 \;
.
\end{eqnarray}
Now, the point of our argument here is that the new vacuum, defined as $b_m |\tilde
0\rangle = 0$ (for all $m$), will not in general be the same as the old vacuum (i.e.
it will not be annihilated by the old annihilation operators, $a_n$). The new vacuum
will, for example, have the following number of particles
\begin{equation}
\langle \tilde 0|a^{\dagger}_i a_i|\tilde 0\rangle = \sum_j
|\beta_{ij}|^2
\end{equation}
with respect to the $i$th mode of $a$ (and this means that the vacuum of the $b$ modes
contains some particles with respect to the $a$ modes). Therefore, since the vacuum
$|\tilde 0\rangle$ is an overall pure state, we can conclude that the amount of
entanglement between the $i$th mode and all the other modes taken together is given by
the von Neumman entropy of entanglement \cite{Vedral}
\begin{equation}
E_{i} = - \sum_j |\beta_{ij}|^2 \ln |\beta_{ij}|^2
\end{equation}
We see that this entanglement is zero only when one of the $\beta$s is a unity and the
rest are zero, i.e. the new modes are the same as the old ones. The maximum
entanglement, on the other hand, is achieved when all the betas are equal. Our two
modes at the beginning of the section are a very special case of this analysis.

In summary, there are two main messages in this section: one is
that entanglement is highly dependent on the way we decide to
divide the whole system into different modes; two is that
interaction in general gives rise to entanglement. There are
numerous examples of various kinds of mode transformations in
quantum optics (for example, in two-mode squeezing), condensed
matter physics as well as relativistic quantum field theory which
lead to entanglement generation as described above. We discuss a
few of them next.

\section{Bose Condensates and The Unruh Effect}

We first briefly review the second quantization treatment of
interacting systems and then apply it to several cases. The
Hamiltonian in the first quantized form is now given by
\begin{equation}
H (x,x') = H_0 (x) + V(x,x')
\end{equation}
where $H_0 (x)$ is the basic (free) Hamiltonian without interactions and the
interaction itself is encoded in the term $V(x,x')$. We already know how to second
quantize $H_0 (x)$, and so we only need to be able to second quantize the interaction
Hamiltonian. We will do this now with the intention of obtaining a typical Hamiltonian
for the Bose-Einstein gas. This is done by performing the following transformation
\begin{equation}
\tilde V = \int dV \Psi^{\dagger} (x, t) \Psi^{\dagger} (x', t)
V(x,x') \Psi (x, t) \Psi (x', t) \; .
\end{equation}
As before, we need to switch from the creation and annihilation
operators in the position picture to the creation and annihilation
operators in the momentum picture. This is performed in the same
way as we did for the non-interacting fields, as in eq.
(\ref{xtop}). After some manipulations and approximations we can
arrive at \cite{Landau}
\begin{equation}
\tilde H = \sum_p \frac{p^2}{2m} a_p^{\dagger} a_p +
\sum_{p_1,p_2,q} \tilde V (q) a_{p_1 + q}^{\dagger} a_{p_2 -
q}^{\dagger} a_{p_2} a_{p_1} \; .
\end{equation}
So, the first term of the Hamiltonian is (as before) a collection of harmonic
oscillators and the second term signifies their interaction. The interaction term
tells us that two particles of momenta $p_1$ and $p_2$ collide to produce new
particles with momenta $p_1+q$ and $p_2 -q$ respectively. The values of the final
momenta are such as to obey the momentum conservation law. $\tilde V(q)$ is the
strength of interaction which is derived from the original potential $V(x,x')$. The
exact form of this is unimportant. In fact, it is sometimes customary to assume that
the interaction is independent of the momentum change $q$ and to write (after omitting
the tilde)
\begin{equation}
H = \sum_p \frac{\hbar p^2}{2m} a_p^{\dagger} a_p + V
\sum_{p_1,p_2,q}  a_{p_1 + q}^{\dagger} a_{p_2 - q}^{\dagger}
a_{p_2} a_{p_1} \; .
\end{equation}
This becomes the same as the Hamiltonian for a degenerate, but
almost ideal Bose gas \cite{Landau}. The creation and
annihilations operators, of course, obey the usual bosonic
commutation relations. In order to solve and find the eigenvalues
of this Hamiltonian we need to convert it into a bunch of
non-interacting harmonic oscillators (or modes). The
transformations that achieve this are called the Bogoliubov
transformations and are given by
\begin{eqnarray}
a_p & = & u_p b_p + v_p b^{\dagger}_{-p} \\
a^{\dagger}_p & = & u_p b^{\dagger}_p + v_p b_{-p}
\end{eqnarray}
where the coefficients $u$ and $v$ (functions of the momentum $p$)
have to be chosen so that the Hamiltonian for $b$ modes has no
interacting parts (see \cite{Ketterle} for an experimental
confirmation of the Bogoliubov transformations in a Bose
condensate). The resulting Hamiltonian, written in terms of the
$b$ modes, is given by
\begin{equation}
H = E_0 + \sum_{p\neq 0} E(p) b^{\dagger}_p b_p
\end{equation}
where $E_0$ is, for us an unimportant, constant and $E(p)$ is the
energy of new modes (all the details, which are irrelevant for the
analysis here, can be found in \cite{Landau}). Here we have a
collection of decoupled (i.e. non-interacting) modes as in the
previous section when we discussed the example of two harmonic
oscillators. Again, the ground state is an entangled state in
terms of the $a$ modes and is given by
\begin{equation}
|\psi_0\rangle = \prod_{p\neq 0} \frac{1}{u_p} \sum_{i=0}^{\infty}
\{-\frac{v_p}{u_p} \}^i |n_{-p} = i ; n_{p} = i\rangle \; .
\end{equation}
In this state only the $p$ and $-p$ modes are entangled and different momenta are not
coupled. The amount of entanglement is again easy to compute to be (c.f. eq.
(\ref{entharm}))
\begin{equation}
E = \sum_{p\neq 0} \{ \frac{\ln (u_p/v_p)^2}{(u_p/v_p)^2 - 1} -
\ln (1 - (v_p/u_p)^2)\}
\end{equation}
This is a sum of entanglements in each individual pair of modes
with the opposite momenta. The entanglement is highest when $u_p
\approx v_p$. This happens in the low momentum regime, more
precisely when $p <<mc$. In the other regime, when the momenta are
high we have that $u_p \approx 1$ and $v_p \approx 0$ and so
entanglement drops to zero.

A very closely related, but even more intriguing, case of mode transformations
generating entanglement can be found in the domain of relativistic quantum field
theory. Next we discuss the accelerated vacuum leading to the Unruh effect
\cite{Davies}. The key point in this effect is that the vacuum state for an uniformly
moving observer in the flat Minkowski space, which is a continuum of decoupled,
disentangled, harmonic oscillators, is actually seen by an accelerated (Rindler)
observer to be entangled. The entanglement is created because the creation operator
for a particle of momentum $p$ for the accelerated observer needs to be expanded as a
sum of the creation operations of the uniform observer involving both the $p$ and $-p$
momenta. More precisely, Minkowski quanta (i.e. those detected by an inertial
observer) are obtained by quantizing the positive frequency solutions of the
Klein-Gordon equation; however, once we change the time parameter which is what
happens for the accelerated Rindler observer, we obtain both the set of positive as
well as the negative frequency solutions. These, when quantized, lead to completely
different set of raising and lowering operators, and therefore a new type of quanta.
Interestingly, the transformation of the mode operators between the stationary and the
accelerated observer is actually of the Bogoliubov type discussed already in the case
of a Bose condensate. The vacuum state of the accelerated observer is, therefore, an
entangled superposition of the number states for $p$ and $-p$ of the original uniform
field \cite{Fulling}. The interesting observation is that the spectrum of the reduced
density matrix (when we trace out the negative momenta for example) is Planckian (just
like in the case of two interacting harmonic oscillators) whose temperature is given
by $T = a/2\pi k$, where $a$ is the acceleration of the observer and $k$ is the
Boltzmann constant. There is, therefore, entanglement between positive and negative
modes due to acceleration, and this entanglement is proportional to the amount of
acceleration. The amount of entanglement, more precisely, is the same as in eq.
(\ref{entharm}) with the appropriate value for the temperature.

We would just like to close this section by briefly mentioning an effect closely
related to the entanglement in the accelerated vacuum and the corresponding thermal
bath. This effect was discovered by Hawking and it explains the radiation from a black
hole and the fact that we can associate a temperature and entropy with it
\cite{Hawking}. The two effects - accelerated observers and the black hole radiation -
are, loosely speaking, related because of Einstein's equivalence principle, which
stipulates that an accelerated observer is the same as an observer in a gravitational
field whose strength is that same as the value of acceleration. So, someone sitting
close to a black hole for example, should also see a radiation (just like an
accelerated observer does) which has a black body spectrum. The temperature of the
radiation can be shown to be $T = \kappa/2\pi k$, where $\kappa$ is the surface
gravity. Hawking \cite{Hawking} showed in $1974$ that this exactly is the case. The
entanglement here is generated because the incoming modes which contain only positive
frequency waves will be converted into the outgoing waves, which will depend also on
the negative frequency waves due to the metric change during the collapse of matter to
form a black hole (there is a pair creation of this type under more general metric
changes in the expanding universe as first discussed by Parker in \cite{Parker}. He
found that particle creation occurs because the positive and negative parts of the
field become mixed during the expansion, such that the creation and annihilation
operators at one time $t$ are a linear (Bogoliubov) combination of those at an earlier
time.). So, the situation here can mathematically be described by the same Bogoliubov
transformation discussed before \cite{Davies,Hawking}. Interestingly, the entropy of a
black hole can also be seen to arise due to entanglement between the modes outside the
black hole and modes inside the black hole \cite{Srednicki} (this is a static, rather
than a dynamic explanation of the entropy of the black hole). There are many issues
involved here which have not been resolved, but which are beyond the motivation of the
current paper, and I direct the interested reader to consult Wald's excellent review
on this and related subjects \cite{Wald}.

The above examples illustrate the entanglement properties in the
second quantization and its dependence of the choice of modes. A
collection of harmonic oscillators that does not interact becomes
interacting when we change to a different set of modes. This
change may be ``passive", like in the case when we diagonalize the
Bose condensate Hamiltonian, or it could be ``active", like when a
stationary observer accelerates. But, the question we would really
like to ask in this paper is whether entanglement could arise
without any interaction between different modes. So far, we have
only considered spinless bosons, and the particle statistics has
not been crucial in any respect. We would now like to analyze the
interplay between the internal and external degrees of freedom in
fermions and bosons, which are otherwise non-interacting.

\section{Non-interacting Fermions and Bosons}

Suppose we have a collection of non-interacting electrons, in a metal, for example.
The electrons all move with various different momenta $p$ and Pauli's exclusion
principle tells us that there can be only two electrons with the same value of
momentum (one with spin, $s$, pointing ``up" and the other with the spin pointing
``down"). The state of these electrons is well approximated by the overall state (the
so called Fermi model of a metal):
\begin{equation}
|\phi_0\rangle = \prod_{s,p} b^{\dagger}_s (p) |0\rangle \; .
\end{equation}
The fact that these are Fermions (as opposed to Bosons) is
reflected in the anticommutation relations which we have to impose
between the annihilation (and creation) operators
\begin{equation}
[b^{\dagger}_s (p), b_{t} (q)]_{+} = \delta_{st} \delta (p-q) \; .
\end{equation}
Obviously there are {\em no} correlations between the momenta of electrons. This means
that if we find an electron and measure its momentum to be $p$, then if we measure
another electron's momentum $q$, its value will be completely independent of the first
measurement. That is exactly what the product of creation operators in the above state
signifies. However, there is a great deal of entanglement in the above state, we just
need to look for it more carefully. Suppose that we measure an electron at the
position $r$ and another one at the position $r'$. We will assume, for the sake of
simplicity, that we have arbitrarily precise detectors (this is not a point of
fundamental importance in the analysis). Suppose further that one electron has spin
up, can we infer the direction of the spin of the other electron? The question is, are
electrons at different locations correlated in in their spin directions? In order to
calculate this, we need to compute the density matrix describing the spin state of two
electrons. It is (by definition) given by
\begin{equation}
\rho_{ss';tt'} = \langle \phi_0| \psi^{\dagger}_{t'} (r')
\psi^{\dagger}_{t} (r) \psi_{s'} (r') \psi_{s} (r) |\phi_0\rangle
\end{equation}
where $\psi^{\dagger}_{t} (r)$ creates a particle of spin $t$ at the location $r$.
This is the same as computing the elements $|st\rangle\langle s't'|$ of the
two-electron density matrix. Assuming (as we usually do) that
\begin{equation}
\psi_s (r) = \int \frac{d^3 p}{(2\pi)^3} e^{ipr} b_s (p)
\end{equation}
we obtain the following form of the density matrix
\begin{eqnarray}
\rho_{ss';tt'} & = & \int \frac{d^3 p}{(2\pi)^3}\frac{d^3
q}{(2\pi)^3}\frac{d^3 p'}{(2\pi)^3}\frac{d^3 q'}{(2\pi)^3}
e^{-i(p-p')r} e^{i(q-q')r'} \times \nonumber\\
& \times & \langle \phi_0| b^{\dagger}_{t'} (p) b^{\dagger}_{t}
(q) b_{s'} (q') b_{s} (p') |\phi_0\rangle \; .
\end{eqnarray}
Now we evaluate the last term to be
\begin{eqnarray}
\langle \phi_0| b^{\dagger}_{t'} (p) b^{\dagger}_{t} (q) b_{s'}
(q') b_{s} (p') |\phi_0\rangle & = &  (2\pi)^6
\delta_{st}\delta_{s't'} \delta (p - q)\delta (p'-q') \nonumber\\
& - & (2\pi)^6 \delta_{s't}\delta_{st'} \delta (p' - q)\delta
(p-q') \; .
\end{eqnarray}
This form is intuitively clear as we have both the direct term, i.e. the term where
the particle with momentum $p$ changes to $q$ and $p'$ to $q'$, and the exchange term,
such that the particle with momentum $p$ changes to $q'$ and $p'$ to $q$. Note that
because of the conservation of momenta these have to be the same (there is no
interaction to allow the change of momenta or conversion of one particle into two for
example). This now enables us to compute the density matrix, by observing that the
density of electrons is given by $n = \int_0^{p_F} dp^3/(2\pi)^3$ (where $p_F$ is the
Fermi momentum),
\begin{equation}
\rho_{ss';tt'} = n^2 (\delta_{st}\delta_{s't'} -
\delta_{s't}\delta_{st'} f^2(r-r')) \; , \label{twoelect}
\end{equation}
where
\begin{equation}
f(r-r') = \int_{0}^{p_F} \frac{d^3 p}{(2\pi)^3} e^{-ip (r-r')} =
\frac{3j_1 (k_F |r-r'|)}{k_F |r-r'|}
\end{equation}
is the ``exchange interaction" term, and  $j_1 (k_F |r-r'|)$ is
the spherical Bessel function with $k_F = p_F/\hbar$. This form of
$\rho$ contains all $16$ elements needed for the spin density
matrix. This $4$ by $4$ density matrix is (up to a normalization)
given by
$$
\rho_{12}=\left(\begin{array}{cccc}
1-f^2&0&0&0\\
0&1&-f^2&0\\
0&-f^2&1&0\\
0&0&0&1-f^2\end{array} \right) \; ,
$$
where the subscript $12$ signifies that there are two electrons, and the basis is the
usual $|\uparrow \uparrow\rangle,|\uparrow \downarrow\rangle, |\downarrow
\uparrow\rangle, |\downarrow \downarrow\rangle$. We first have to decide if this
matrix is entangled. We perform the partial transposition and compute the resulting
eigenvalues \cite{Horodecki}. They are $1-f^2$ (doubly degenerate), $1$ and $1-2f^2$.
The first two eigenvalues are positive as $0\le f\le 1$, but the last one can be
negative. The condition is that $f^2>1/2$, and in this region we have entanglement.
Therefore the density matrix is entangled for the region of distances $0\le |r-r'| <
r_e$, where $r_e $ is the value such that $j^2_1 (k_F r_e) =1/2$ (the value is about
$r_e \approx \pi \hbar/8 p_F$ as expressed in terms of Fermi's momentum).

We can compute the entanglement of formation \cite{Wootters} of the spins for a
general state $\rho_{12}$ (as a function of $f$). It is given by
\begin{eqnarray}
E (\rho_{12}) & = &-\frac{1+\sqrt{1-(\frac{2f^2-1}{2-f^2})^2}}{2} \ln
\frac{1+\sqrt{1-(\frac{2f^2-1}{2-f^2})^2}}{2} \nonumber \\
& - & \frac{1+\sqrt{1+(\frac{2f^2-1}{2-f^2})^2}}{2} \ln
\frac{1+\sqrt{1+(\frac{2f^2-1}{2-f^2})^2}}{2}
\end{eqnarray}
Entanglement is maximal when $f=1$ and it is equal to $\ln 2$. The resulting state is
the spin singlet state $|\uparrow\downarrow\rangle - |\downarrow\uparrow\rangle$.
Otherwise, as the distance decreases and $f$ drops, the more of the triplet state
becomes mixed in with the singlet and entanglement drops as a consequence. Beyond the
distances of about $\pi \hbar/8 p_F$, all entanglement disappears. A reasonable
estimate for the Fermi momentum in a metal is $10^{-27} kg 10^4 ms^{-1}$, which gives
a distance of about $\Delta r > 10^{-7}$ meters beyond which no two electrons are
entangled in spin. Below this distance, every two electrons in a conductor are
entangled. If we have a much bigger momentum, such as in a neutron star, then the
entanglement distance becomes very small. For a typical neutron star, this distance
would be about $10^{-15}$ meters, which tells us that there is no entanglement there
between any two neutrons in a neutron star. Therefore, the higher the Fermi momentum
the smaller the ``entanglement distance" between the electrons. Note that it is a
separate issue whether this entanglement present in the spin degrees of freedom can
actually be used to process information, such as in the teleportation protocol. In
order to teleport, for example, the electrons need to be separated sufficiently well
after being detected. How feasible this is in practice will be a subject of a separate
investigation.

The disappearance of entanglement is intuitively expected from the following simple
argument. Quantum statistics becomes important only when the spatial wavefunctions of
individual fermions have a substantial overlap. The wavefunction spread of fermions is
roughly given by the wavelength which can be estimated to be $\hbar /p_F$. Therefore
beyond this distance we expect particle statistics not to play such a strong role. Let
us look at this in the first quantization to make the case even clearer. Suppose that
the total (un-normalized) state of two electrons is:
\begin{equation}
|01\rangle |\psi_1,\psi_2\rangle
\end{equation}
where the $|\psi_1,\psi_2\rangle$ are the spatial wavefunctions of
the two electrons, and $0,1$ are the values of spin (the first
position always belongs to the first electron and the second to
the second electron). Suppose that $\langle \psi_1 | \psi_2
\rangle =\epsilon$. Now this is not an allowed state if the
particles are truly indistinguishable and in the case of fermions
it needs to be antisymmetrized. The corresponding allowed state is
\begin{equation}
|01\rangle |\psi_1,\psi_2\rangle - |10\rangle
|\psi_2,\psi_1\rangle
\end{equation}
Now, this state can be entangled in spin. To check this, we trace
out the spatial degree of freedom to obtain the following density
matrix (again unnormalised)
\begin{equation}
|01\rangle\langle 01| + |10\rangle\langle 10| - \epsilon^2
(|01\rangle\langle 10| + |10\rangle\langle 01|)
\end{equation}
It can be checked, again using the partial transposition
criterion, that this state is entangled, unless $\epsilon = 0$,
which means that the two electrons have no spatial overlap.
Therefore we see that the quantum statistics plays a crucial role
for entanglement.

Classical correlations, on the other hand, are still present even beyond this
distance. This is true in the above first quantized model, as well as in the model of
many non-interacting fermions. We can quantify them using the mutual information,
$I(\rho_{12}) = S(\rho_1) + S(\rho_2) - S(\rho_{12})$, in the state in eq.
(\ref{twoelect}) \cite{Vedral}. Here $\rho_1$ and $\rho_2$ are the reduced density
matrices derived from $\rho_{12}$. Computing this for our state of two electrons we
obtain
\begin{eqnarray}
I(\rho_{12}) & = & 2 \ln 2 + 2 \frac{1-f^2}{4-2f^2} \ln
\frac{1-f^2}{4-2f^2} \nonumber\\
& + & \frac{1}{4-2f^2}\ln \frac{1}{4-2f^2} + \frac{1-2f^2}{4-2f^2}
\ln \frac{1-2f^2}{4-2f^2}
\end{eqnarray}
We see that the classical correlations only vanish when $f=0$ as we then have an equal
mixture of all four states $|\uparrow, \uparrow \rangle, |\uparrow, \downarrow
\rangle, |\downarrow, \uparrow \rangle, |\downarrow, \downarrow \rangle$. Otherwise,
spins are always (classically) correlated.

The entanglement in spin when space is traced out is of a similar
(but somewhat dual) nature to the spin-to-space entanglement
transfer in the double beam-splitter set up in \cite{Omar}. The
entanglement, which in this experiment by Omar el. al. initially
exists in the spin degrees of freedom, is, due to the quantum
statistical nature of particle interaction, transferred to space.
This also is a purely quantum statistical effect as there is no
other interaction between the internal and external degrees of
freedom (although there is an interaction between the spatial
degrees of freedom of different system involved).

Note that the presented method here allows us to extract more complicated information
such as correlations between three and more electrons. These would be increasingly
more difficult to treat mathematically, but the calculations can be performed in
principle. We could, for example, check if genuine multi-electron correlations exist
and what are the relevant distances to allow such correlations. Let us briefly
summarize the calculation: we have three electrons at positions $r,r'r''$ and they
have some values of spin. We would like to extract the spin density matrix, which is
now at $8$ by $8$ square matrix. The form of the matrix can actually be deduced
without much effort from the simple ``pictorial" representation. We have the following
terms
$$
\rho_3=\left(\begin{array}{ccc}
s&s'&s''\\
t&t'&t''\end{array} \right) -
\left(\begin{array}{ccc}
s&s'&s''\\
t&t''&t'\end{array} \right)-
\left(\begin{array}{ccc}
s&s'&s''\\
t'&t&t''\end{array} \right)-
\left(\begin{array}{ccc}
s&s'&s''\\
t''&t'&t\end{array} \right)+
$$
$$
\left(\begin{array}{ccc}
s&s'&s''\\
t'&t''&t\end{array} \right)+
\left(\begin{array}{ccc}
s&s'&s''\\
t''&t&t'\end{array} \right)
$$
There are six terms as there are six possible arrangements; every spin in the density
matrix can initially be paired with any of the three spins. The next spin has only two
possibilities and the last one has only one remaining choice. The choice of signs
should also be clear: the first term is the direct term and therefore positive. The
next three terms involve one double exchange and are therefore negative, and the last
two terms involve triple exchanges and are therefore positive. The density is,
therefore, given by
\begin{eqnarray}
\rho (s,s',s'';t,t',t'') & = & \langle \phi_0|\psi^{\dagger}_{t''}
(r'') \psi^{\dagger}_{t'} (r') \psi^{\dagger}_{t} (r) \psi_{s}
(r') \psi_{s'} (r) \psi_{s''} (r'')
|\phi_0\rangle \nonumber \\
& + & n^3 (\delta_{st}\delta_{s't'}\delta{s''t''} \nonumber \\
& - & f_1 f_2 \delta_{st}\delta_{s't''}\delta{s't''} \nonumber \\
& - & f_1 f_3 \delta_{st'}\delta_{s't}\delta{s''t''} \nonumber \\
& - & f_2 f_3 \delta_{st''}\delta_{s't'}\delta{s''t} \nonumber \\
& + & f_1 f_2 f_3 \delta_{st'}\delta_{s't''}\delta{s''t} \nonumber \\
& + & f_1 f_2 f_3 \delta_{st''}\delta_{s't}\delta{s''t'})
\end{eqnarray}
There are three different $f$ functions now as we have three arguments $|r-r'|$,
$|r-r'|$ and $|r'-r''|$.  This contains information about all $64$ elements. The
actual matrix is (for simplicity in notation, I will assume that all the $f$ functions
are equal)
$$
\rho_{123}=\left(\begin{array}{cccccccc}
1-3f^2+2f^3&0&0&0&0&0&0&0\\
0&1-f^2&-f^2&0&-f^2+f^3&0&0&0\\
0&-f^2&1-f^2&0&-f^2+f^3&0&0&0\\
0&0&0&1-f^2&0&-f^2+f^3&-f^2+f^3&0\\
0&-f^2+f^3&-f^2+f^3&0&1-f^2&0&0&0\\
0&0&0&-f^2+f^3&0&1-f^2&0&0\\
0&0&0&-f^2+f^3&0&0&1-f^2&0\\
0&0&0&0&0&0&0&1-3f^2 + 2f^3\end{array} \right),
$$
In order to confirm that we obtain the same $2$ electron density
matrix, we have to trace one electron out, and this we have to do
by tracing over both the spin and space. If this is performed, we
can easily confirm that the resulting density matrix is the one
obtained previously for two electrons.

Now, we would like to know if this state contains three party entanglement in spin. We
already know that every two parties can be entangled in the appropriate distance
region, for example, when they are very close to each other. What happens when
$r=r'=r''$? Then, $f_i = 1$ and we see that the resulting state is actually not a
physical state (all the elements go to zero indicating that this is not a physical
state - or that the probability for this to happen vanishes). This is physically
clear, as when we have three spins, at least two out of three have to be in the same
state and this is not possible as they are all in the same physical location. At the
other extreme, when all three electrons are very far away, then the density matrix
again becomes maximally mixed. There is no entanglement in this case of any kind, nor
do we have classical correlations. We also have a number of other possibilities, such
as, for example, that two electrons are close to each other, but the third one is
further away from them. We could then imagine that all three are in an entangled state
as well as having entanglement between the two close by electrons. Let us take as an
example the state when all electrons are very close to each other. Then, to a good
approximation, the state is an equal mixture of the four state $|000\rangle$,
$|111\rangle$, $|001\rangle + |010\rangle + |100\rangle$ and $|011\rangle +
|110\rangle + |101\rangle$. Now, this state clearly contains three party entanglement.
This is because every electron is entangled to the other two electrons, indicating
that the state cannot be separable in any way. Following this logic therefore we have
a three party entanglement providing $f^2 > 1/2$. Can we have a genuine three part
entanglement beyond this point even though there is no two party entanglement? This
would be something like a GHZ state (where there is no two party entanglement, but all
three parties are entangled) and this remains an open question. Generalizing our line
of reasoning, however, we can state that if a bunch of $n$ electrons is found within
the distance of roughly the inverse Fermi wavenumber, then the corresponding total
state of their spins would be an entangled state on all levels (two, three-party and
so on up to $n$-party entanglement). The above treatment also allows us to compute
various mutual informations between different parties involved.

Interestingly, an equivalent analysis can be performed for a bunch of non-interacting
bosons (with polarization as the internal degree, for example), but we will not see
any correlations in polarization. This is because there is no restriction on how many
bosons can be in a state with a given momentum and polarization. The most general
state of bosons is now given by
\begin{equation}
|\phi\rangle \propto \prod_{sp} (b_s^{\dagger} (p))^{n(p,s)}
|0\rangle
\end{equation}
Where $n(p,s)$ is the number of bosons of momentum $p$ and
polarization $s$. The operators obey the following commutation
relations
\begin{equation}
[b^{\dagger}_s (p), b_{t} (q)]_{-} = \delta_{st} \delta (p-q)
\end{equation}
It is now intuitive that if we measure a photon of one polarization at $r$, there is
no reason why this should be correlated to the polarization of another photon at $r'$.
We can compute the same density matrix as in the case of fermions, but the quantity
$\langle \phi_0| b^{\dagger}_{t'} (p) b^{\dagger}_{t} (q) b_{s'} (q') b_{s} (p')
|\phi_0\rangle$ will now be independent of the labels $s,s',t,t'$. This means that the
resulting $4$ by $4$ density matrix of polarization will have all the $16$ elements
equal to each other. This represents a disentangled pure state $(|H\rangle +
|V\rangle)\otimes (|H\rangle + |V\rangle)$, where $H$ and $V$ are the horizontal and
vertical polarizations respectively (the polarization plays the same role here as the
spin did in the case of fermions). And, so, here we see a fundamental difference
between fermions and bosons with respect to the interplay between the entanglement in
internal and spatial degrees of freedom. Therefore, there is no correlation between
the relative distance between the photons and their polarization. Bosons, in other
words, do not obey the Pauli exclusion principle.

Note that in this section we have analyzed the presence of entanglement in the
internal degrees of freedom and have not talked about the spatial degrees of freedom.
The analysis of long range correlations in fermions and bosons can be found in
\cite{Yang}, although the author there does not distinguish between classical and
quantum correlations (and the internal degrees of freedom play no special part).
Nevertheless, similar methods could possibly be used in the future to perform a more
detailed analysis of more complicated systems. Note also that all our discussions here
could be generalized to overall mixed state (here the total states we start with are
pure states) as in the statistical mechanics of many body systems \cite{Lee}. The same
analysis can be performed as the one presented above to calculate the amount of
entanglement in the internal degrees of freedom. It is to be expected, however, that
this additional mixedness can only reduce the existing correlations in an overall pure
state computed here. Otherwise, the results would not be fundamentally different to
the ones presented here.

\section{Concluding Remarks}

We have seen that the second quantization formalism adds some interesting effects to
our understanding of entanglement. First and foremost, this is because we have to
decide what our subsystems are before we can speak about the amount of entanglement
between them. This is not an issue in the first quantized version of quantum mechanics
that is used when we deal with a small number of systems, such as $2$ or $3$ (by
definition) distinguishable) qubits. An interesting observation is that the change of
modes also affects the amount of entanglement present. A change of modes is, of
course, a ``global change" relating the old modes to the new ones and this is why
entanglement can be affected in the first place - ``local changes" on their own would
not be able to generate entanglement (We have to be a bit more careful here as we have
to restrict ourselves to speaking about modes and excitations of the modes. We cannot
really speak about local operations on ``particles", since they are identical and
whatever operation we perform on one of them, we, in fact, have to perform on all of
them). The change of modes produces an interaction between the new modes which is what
gives rise to entanglement. In section $3$, we presented the most general formalism
for mode changes and showed how the amount of entanglement varies with the change of
modes. However, we have also been able to show that we do not really need an
interaction to produce entanglement. We can instead rely on the effects of quantum
particle statistics to produce entanglement. This effect was illustrated by looking at
the Fermi sea model of a conductor, where we had a collection of non-interacting
electrons (neither in spin nor in space) occupying states up to the highest Fermi
level (assumed arbitrary). Here we found that two electrons are entangled in the spin
degrees of freedom providing that their distance is (roughly) below the inverse Fermi
momentum (measured in units of $\hbar =1$). Our method can be used to compute the
density matrix for any number of subsystems. We have also pointed out a crucial
difference between non-interacting fermions and non-interacting bosons, in that
non-interacting fermions can still be found to be entangled in the internal degrees of
freedom, while bosons cannot. It would be interesting to investigate the further
possibility of using these ``non-interacting" correlations to perform information
processing \cite{Paunkovic}. It would also be interesting to extend our discussion to
more complicated systems composed of fermions and bosons.

\vspace*{0.5cm}

\noindent {\bf Acknowledgments.} I would like to thank A. J. Fisher, J. R. Gittings,
N. Paunkovi\'c, J. Pachos and G. Castagnoli on many interesting and helpful
discussions on the nature of the link between entanglement and particle statistics. S.
J. van Enk and J. Schliemann are acknowledged for pointing out some related references
to this work. My research on this subject is supported by European Community,
Engineering and Physical Sciences Research Council, Hewlett-Packard and Elsag spa.

\end{document}